%% file: paper.tex
\documentclass[]{aa}
\usepackage{graphicx}
\usepackage[utf8]{inputenc}
\usepackage[T1]{fontenc}
\usepackage{epstopdf}
\usepackage{cuted}
\usepackage{caption}
\usepackage[bookmarks=false]{hyperref} 
\usepackage{amsmath}
\usepackage{subcaption}
\usepackage{graphicx}
\usepackage{verbatim}
\usepackage{natbib}
\usepackage{rotate}
\usepackage{dsfont}
\usepackage{lscape}
\usepackage{aalongtable}
\usepackage{supertabular}
\usepackage{mathbbol}
\usepackage{bm}
\usepackage{mathtools}
\usepackage{enumerate}
\usepackage{amsmath}
\usepackage{MnSymbol}
\allowdisplaybreaks
\usepackage{float}

\bibpunct{(}{)}{;}{a}{}{,}

\newcommand{\appropto}{\mathrel{\vcenter{
  \offinterlineskip\halign{\hfil$##$\cr
    \propto\cr\noalign{\kern2pt}\sim\cr\noalign{\kern-2pt}}}}}

\begin{document}

\title{On the variance of radio interferometric calibration solutions}
\subtitle{Quality-based Weighting Schemes}

\author{Etienne Bonnassieux$^{1,2}$, Cyril Tasse$^{3}$, Oleg Smirnov$^{2,5}$, Philippe Zarka$^{1,4}$} 

\institute{
LESIA, Observatoire de Paris, PSL, CNRS, Sorbonne Universités, UPMC Univ. Paris 06, Univ. Paris Diderot, Sorbonne Paris Cité, 5 place Jules Janssen, 92195 Meudon, France
\and
Department of Physics \& Electronics, Rhodes University, PO Box 94,
Grahamstown, 6140, South Africa
\and
GEPI, Observatoire de Paris,
5 place Jules Janssen, 92190 Meudon, France
\and
Station de Radioastronomie de Nançay (SRN), Observatoire de Paris, CNRS USN, PSL Research University, Université d'Orléans, 18330 Nançay, France
\and
South Africa Radio Astronomy Observatory, 3rd floor, The Park, Park Roads 7405 Pinelands, South Africa
}

\authorrunning{E. Bonnassieux, C. Tasse, O. Smirnov, P. Zarka}

\input{source/Defs.tex}

\input{source/abstract.tex}

\titlerunning{On the variance of radio interferometric calibration solutions}
   \maketitle

%
%
%

\input{source/intro.tex}

\input{source/matrix-formalism.tex}

\input{source/matrix-noise-map-simulations.tex}

\input{source/flattening.tex}


\input{source/matrix-algorithm.tex}
\input{source/discussion.tex}

\begin{acknowledgements}
This work is based
upon research supported by the South African Research Chairs Initiative of the Department of Science and Technology and National Research Foundation. The authors thank the LOFAR Surveys KSP Primary Investigators for allowing us to use their data for our tests and images in this paper. They would also like to thank M. Atemkeng and T. Gobler for fruitful discussions over the course of this work. They helped shape this work through their insightful comments on both style and substance while proofreading.
\end{acknowledgements}

\bibliographystyle{aa}
\bibliography{source/bib}




\end{document}

%% file: source/Defs.tex

\def\pg{ }

\newcommand\bydefn{\hspace{3pt}\stackrel{\mathclap{\mbox{\tiny{def}}}}{=}\hspace{3pt}}
\newcommand\approxdefn{\hspace{3pt}\stackrel{\mathclap{\mbox{\tiny{def}}}}{\approx}\hspace{3pt}}
\def\estC{\widehat{\bm{C}}_{\matGains}}
\def\sdnu{s_d^\nu}
\def\tnuIndeix{\tau}
\def\Noisefunc{\mathcal{N}_{\tnuIndeix\tnuIndeix'}}
\def\Kfunc{\mathcal{K}_{\tnuIndeix\tnuIndeix'}}
\def\Errfunc#1{\mathcal{E}_{#1}}
\def\deltu{\delta u}
\def\deltv{\delta v}
\def\deltl{\delta l}
\def\deltm{\delta m}

\def\weight#1{\omega_{pq,#1}}
\def\deltasamp#1{\delta\left(#1-\underline{t}_0\right)}
\def\Sampfunc#1#2{\mathcal{C}_{#1}\left(#2\right)}
\def\DirFunc#1{\phi_{#1}}
\def\FluxFunc#1{\mathcal{S}_{#1}}
\def\WeightFunc#1{\mathcal{W}_{#1}}
\def\dudv{\delta u \delta v}
\def\t0{\mathbf{\tnuIndeix_0}}
\def\Rpqt{\tilde{r}_{pq}^\tnuIndeix}
\def\CovMat{\bm{C}}
\def\Order#1{\mathcal{O}\big(#1\big)}
\def\Apqtlm{\mathcal{A}_{pq,\tnuIndeix}^{lm}}
\def\Apqtlmprime{\mathcal{A}_{pq,\tnuIndeix'}^{lm}}
\def\Spqt{S_{pq,\tnuIndeix}}
\def\Spqtprime{S_{pq,\tnuIndeix'}}
\def\diag#1{\text{Diag}\left\{#1\right\}}
\def\noise{\bm{n}}

\def\gvect{\textbf{g}}
\def\ghatvect{\hat{\textbf{g}}}
\def\Ghatvect{\hat{\textbf{G}}}

\def\Var#1{\textrm{Var}\{#1\}}
\def\Cov#1{\textrm{Cov}\{#1\}}
\def\Exp#1{\textrm{E}\{#1\}}
\def\tExp#1{\textrm{E}_{\mathcal{C}}\{#1\}}
\def\Norm#1{\|#1\|}
\def\hadam{\circ}

\def\schur{\circ}

\def\dblW{\bm{\Omega}}
\def\dblS{\bm{\beta}}
\def\dblF{\bm{\Gamma}}
\def\matGains{\bm{\tilde{\gamma}}}
\def\vecw{\bm{w}}

\def\matVt{\bm{{V}}_{pq}^{t\nu}}
\def\matk{\bm{k}}
\def\muelC{\bm{\mathcal{C}}}
\def\muelV{\bm{{V}}}
\def\muelVr{\bm{{\tilde{V}}}}
\def\muelK{\bm{\mathcal{K}}}
\def\muelS{\bm{\mathcal{S}}}
\def\matS{\bm{S}}
\def\matW{\bm{W}}
\def\matw{\bm{w}}
\def\matVtau{\bm{{V}}_{pq}^{\tnuIndeix}}
\def\matVdt{\bm{{V}}_{d,\tnuIndeix}}
\def\matJ#1{\bm{{J}}_{#1,t\nu}}
\def\opvec#1{\mathrm{vec}\{#1\}}
\def\muelM{\bm{\mathcal{M}}}
\def\muely{\tilde{\bm{{{y}}}}}
\def\muelyr{\bm{{\tilde{y}}}}
\def\1{\bm{\mathbb{1}}}
\def\muelF{\bm{\mathcal{F}}}
\def\muelW{\bm{\mathcal{W}}}
\def\hatmuelM{\bm{\mathcal{\hat{M}}}}
\def\muelMr{\bm{\mathcal{\tilde{M}}}}
\def\Bmat#1{\bm{{B}}_{#1}}
\def\muelB{\bm{\mathcal{B}}}
\def\matKtd{\textbf{K}_{d,\tnuIndeix}}
\def\matVV{\bigdoublevee}
\def\matR{\bm{R}}
\def\matRt{\bm{R}_\tnuIndeix}
\def\matRdt{\bm{R}_{d,\tnuIndeix}}
\def\const{\mathrm{const.}}

\def\Rmat#1{\bm{r}_{#1}}
\def\Kmat#1{\bm{K}_{#1}}
\def\Varmat{\widehat{\bm{V}_{\ghatvect}}}
\def\VarmatTau{\widehat{\bm{V}}_{\ghatvect_\tnuIndeix}}
\def\VarmatC{\widehat{\bm{V}_{\cellindex}}}
\def\VarmatCprime{\widehat{\bm{V}_{\cellindex'}}}
\def\VarmatTauprime{\widehat{\bm{V}}_{\ghatvect_{\tnuIndeix'}}}
\def\ExpVarmat{\bm{V}_{\ghatvect_\tnuIndeix}}
\def\kappamat#1{\bm{\kappa}_{#1}}
\def\cellindex{\mathcal{C}}
\def\gptnu{g_p^{t\nu}}
\def\gqtnu{g_q^{t\nu}}

\def\vecktd{\textbf{k}_{d,\tnuIndeix}}
\def\vecg{\vec{g}}
\def\VecHatg{\widehat{\textbf{g}}}
\def\VecHatgt{\widehat{\textbf{g}_\tnuIndeix}}
\def\VecHatgO{\widehat{\textbf{g}_{0}}}
\def\VarVecHatg{\textbf{V}_{\widehat{\textbf{g}}}}

\def\kbd#1{\kappa_{#1}^d}
\def\sqrts{\sqrt{s_d}}
\def\gpt{g_p^{\tnuIndeix}}
\def\gjt{g_j^{\tnuIndeix}}
\def\gqt{g_q^{\tnuIndeix}}
\def\gdjt{g_j^{\tnuIndeix'}}
\def\gdqt{g_q^{\tnuIndeix'}}
\def\Hatgpt{\widehat{g_p^{\tnuIndeix}}}
\def\Hatgp{\widehat{g_p}}
\def\Hatgqt{\widehat{g_q^{\tnuIndeix}}}
\def\Hatgdqt{\widehat{g_q^{\tnuIndeix'}}}
\def\Hatgdpt{\widehat{g_p^{\tnuIndeix'}}}
\def\Hatgjt{\widehat{g_j^{\tnuIndeix}}}
\def\Hatgdjt{\widehat{g_j^{\tnuIndeix'}}}
\def\vpqt{v_{pq}^{\tnuIndeix}}
\def\kpqdt{k_{pq,\tnuIndeix}^{d}}
\def\kpqdtprime{k_{pq,\tnuIndeix'}^{d}}
\def\kpdt{k_{p,\tnuIndeix}^{d}}
\def\kpdtprime{k_{p,\tnuIndeix'}^{d}}
\def\kqdt{k_{q,\tnuIndeix}^{d}}
\def\kqdtprime{k_{q,\tnuIndeix'}^{d}}
\def\kpqtlm{k^{lm}_{pq,\tnuIndeix}}
\def\kpqtlmprime{k^{lm}_{pq,\tnuIndeix'}}
\def\sd{s_{d}}
\def\Irpqlm{I^r_{pq,lm}}
\def\Irpqlmprime{I^r_{p'q',lm}}
\def\Vrpqt{\widetilde{\gamma}_{pq,\tnuIndeix}}
\def\Vrpjt{\widetilde{g}_{pj,\tnuIndeix}}
\def\Vrpqtprime{\widetilde{\gamma}_{pq,\tnuIndeix'}}
\def\Vrjqtprime{\widetilde{\gamma}_{jq,\tnuIndeix'}}
\def\I{\bm{\mathrm{I}}}
\def\J{\bm{\mathrm{J}_2}}
\def\kvect#1{\bm{k}_{#1}}

\def\COH{{\sc CohJones}}

\def\u{u}
\def\v{v}
\def\w{w}
\def\l{l}
\def\m{m}
\def\n{n}
\def\d{d}
\def\dbf{{\bm d}}

\newcommand{\conj}[1]{\overline{#1}}
\newcommand{\conjp}[1]{\left({\overline{#1}}\right)}
\def\vec#1{\ensuremath{\mathbf{#1}}}

\def\JMat{\textbf{J}}
\def\Skyd{\textbf{S}_d}
\def\Vbl{\textbf{V}_{\left(pq\right)t\nu}}

\newcommand{\mat}[1]{{\bm{#1}}}
\newcommand{\JJ}{\mat{J}} 
\newcommand{\JHJ}{\JJ^H\JJ} 
\newcommand{\UL}{\mathrm{UL}}
\newcommand{\Na}{N_\mathrm{ant}}
\newcommand{\Nbl}{N_\mathrm{bl}}
\newcommand{\Nd}{N_\mathrm{dir}}

\def\vbltnu{\textbf{v}_{\left(pq\right)t\nu}}
\def\vbl{\textbf{v}_{pq}}
\def\V{\textbf{V}}
\def\gwirt{\bm{g_{\!_{W}}}}
\def\dgwirt{\dbf\bm{g_{\!_{W}}}}
\def\g{\bm{g}}
\def\vis{\textbf{v}}
\def\Vmeas{\textbf{P}}

\def\separator{
\hrule
\begin{center}
\textsc{to be modified after that}
\end{center}
\hrule
}

\def\Kron{\otimes}

\def\SimpleJacob{\bm{J}}
\def\Jacob{\bm{\mathcal{J}}}
\def\JVpq{\Jacob_{{\textbf{v}_{pq}},\bm{g_{\!_{W}}}}}

\def\JVpqg{\Jacob_{{\textbf{v}_{pq}},\bm{g}}}
\def\JVpqCg{\Jacob_{{\textbf{v}_{pq}},\bm{\conj{g}}}}
\def\JVAtg{\JV\big|_{\vec{g}}}

\def\A{\textbf{A}}
\def\H{\textbf{H}}

\def\JV{\Jacob\left\{\textbf{v}\right\}}
\def\JV{\Jacob_{\textbf{v}}}
\def\JVg{\Jacob_{\textbf{v},\bm{g}}}
\def\JVCg{\Jacob_{\textbf{v},\bm{\conj{g}}}}

%% file: source/abstract.tex
\abstract{
This paper investigates the possibility of improving radio interferometric images using an algorithm inspired by an optical method known as ``lucky imaging", which would give more weight to the best-calibrated visibilities used to make a given image.
A fundamental relationship between the statistics of interferometric calibration {solution residuals} and those of the image-plane pixels is derived in this paper. This relationship allows us to understand and describe the statistical properties of the residual image.
In this framework, the noise-map can be described as the Fourier transform of the covariance between residual visibilities in a new {differential Fourier plane}. 
Image-plane artefacts can be seen as \emph{one realisation} of the pixel covariance distribution, which can be estimated from the antenna gain statistics.

Based on this relationship, we propose a means  of improving images made with calibrated visibilities using weighting schemes.
This improvement would occur after calibration, but before imaging - it is thus ideally used between major iterations of self-calibration loops.
Applying the weighting scheme to simulated data improves the noise level in the final image at negligible computational cost.
}

%% file: source/intro.tex
\section{Introduction}\label{sec.intro}

\pg
Interferometers sample Fourier modes of the sky brightness distribution corrupted by instrumental and atmospheric effects rather than measuring the sky brightness directly. 
This introduces two problems for astronomers to invert: calibration and imaging. Both of these problems are ill-conditioned.

\pg
The problem of imaging consists of correcting for the incomplete $uv$-coverage of any given interferometer by deconvolving the instrument's {Point-Spread Function (PSF)} from images. Its poor conditioning comes from our limited a priori knowledge of the sky brightness distribution, combined with large gaps in our $uv$-coverage, which prevents us from placing strong constraints on image deconvolution. It can be better-conditioned in different ways, including through the use of weighting schemes \citep[see][and references therein]{1995AAS...18711202B,2014MNRAS.444..790Y} to improve image fidelity at the start of deconvolution. When inverting the imaging problem, we often assume that {the sky} is stable within the domain (i.e. {is} constant in time and frequency). There are exceptions, such as wide-band deconvolution algorithms \citep[e.g.][]{2011A&A...532A..71R} that explicitly take into account the {sky}'s frequency-dependence, but still assume that the sky brightness distribution does not vary with time.

\pg
The problem of calibration is what concerns us in this paper. It consists of estimating and correcting for instrumental errors (which includes effects such as antenna pointing errors, but also the phase-delays caused by ionospheric activity, troposphere, etc). 
 {Calibration consists of solving for gain estimates, where a gain models the relationship between the electromagnetic field of an astrophysical source and the voltage that an antenna measures for this source. Because measurements are noisy, calibration often involves some fine-tuning of solution intervals, to ensure that the solutions are well-constrained while the solution intervals stay as small as signal-to-noise allows.} The calibration inverse problem involves three competing statistical effects: thermal noise in the measurements, true gain variability, and {sky} model incompleteness. If gain solutions {are sought individually for each measurement}, then calibration estimates will be dominated by thermal noise, and will not adequately describe the actual gains. Similarly, if a single gain estimate is fitted to too many measurements, the intrinsic gain variability will be ``averaged out''; for example, a choice of time and frequency interval that is too large will cause the solver to estimate a constant gain while the underlying function varies quickly, thereby missing much of the gain structure. This will introduce error which will be correlated in time and frequency. This occurs, for example, when solving for ionospheric phase delays: in the most extreme case, where the solution interval is significantly larger than the scale of ionospheric fluctuations, its varying phase can average out to zero over the interval in time and frequency. Finally, if the model being fitted is incomplete, unmodeled physical flux will likely be absorbed unpredictably into both the gain solutions and the residual visibilities: this absorption of physical flux into gain solutions is known as source suppression \citep[see][and references therein]{2014MNRAS.439.4030G,2013MNRAS.435..597K}.

\pg
In practice, it is reasonable to assume that gain variation is generally slower than some given scale: we can then reduce the noise of our gain estimates by finding a single gain solution for a small number of measurements, assuming that the underlying gain variation is very small and stable over short intervals. This is \emph{generally} a valid hypothesis, but the specific value for the variation scale can be contentious. Indeed, while the noise level can often be treated as constant throughout an observation, the gain variability itself is generally not constant: there will be time periods where the gains will tend to remain constant for longer, and others where variability will be very quick. This means that, for any choice of calibration interval, some gain estimates will be better than others, and almost all could have been improved (at a cost to others) by a different choice of time (and frequency) intervals. 

\pg
Since we have measurements which are better-calibrated than others (in that better estimates for their gains were obtained through chance alone), we could, in principle, take inspiration from ``lucky imaging'' \citep[an optical-domain method for making good images: for more details, see][and references therein]{1978JOSA...68.1651F} to weigh our visibilities according to their calibration quality. 
{Those weights would in effect be an improvement of currently existing methods such as clipping noisy residual visibilities: in the extreme case where all visibilities are equally-well calibrated except a few which are extremely noisy, it should be equivalent to clipping. Otherwise, the weights should show at least a slight improvement over clipping.}

\pg
The key finding of the present paper is a fundamental relationship between the covariance matrices of residual visibilities and the map of the covariance in the image-plane: the ``Cov-Cov relationship'' between visibility covariance to image-plane covariance. We show that the pixel statistics in the image-plane are determined by a ``noise-PSF'', convolved with each source in the sky (modeled or not). This noise-PSF is the product of the Fourier transforms of the gain covariance matrix with each cell mapped not from $uv$ space to $lm$ coordinates but rather between their respective covariance spaces - from a new {differential Fourier plane (henceforth ``$\left(\deltu\deltv\right)$''-plane) to the image-plane covariance space $\deltl\deltm$. This image-plane covariance space describes} the variance in each pixel and the covariance between pixels\footnote{{The noise-PSF also relates $\delta w$ to $\delta n$, as shown in the matrix formalism, but this is not explicitly referenced in the text since visibility space is usually referred to as ``the UV-plane'' in literature, rather than ``the UVW-space''.}}. It describes the expected calibration artefacts and thermal noise around each source, does not vary as a function of direction, and is convolved with each source in the field to yield the final error map. Because all unwanted (in our case, unphysical) signal can be thought of as noise, we will refer to {the pixel variance map} as the ``noise-map''. 

\pg
The notion of a $\left(\deltu\deltv\right)$-plane arises organically from the framework of radio interferometry: we are associating a \emph{correlation} between visibilities to \emph{coordinates} in covariance space, {just as we associate} the visibilities themselves to the $uv$-domain. The $\left(\deltu\deltv\right)$-plane is the natural domain of these correlations. As previously stated, even if all sources in the field are perfectly known and modeled, a poor choice of calibration interval can introduce correlated noise in the residuals, which would then introduce larger variance near sources in the noise-map. Conversely, if calibration is perfect, the noise-map should be completely flat {(i.e. same variance for all pixels)}, as there would be no noise-correlation between pixels. 

\pg
The main result of this paper consists of describing a new adaptive, quality-based weighting scheme based on this insight. Using the Cov-Cov relationship, we can {create a new} weighting scheme by estimating the residual visibility covariance matrix in a given observation. By weighting {visibilities} so as to change their covariance matrix, one can change the shape of the noise-PSF and thus improve {the final image}: this manifests as either decreased noise or decreased calibration artefacts. Note that this weighting is applied after calibration, but before image deconvolution: applying it will therefore not only improve the residual noise in the image, and thus the sensitivity achievable with a given pipeline, but will also improve deconvolution by minimising calibration artefacts in the field: it should thus effectively remove spurious, unphysical emission from final data products. Estimating the covariance matrix is the main difficulty of our framework: we do not know the underlying covariance matrix, and the conditioning of our estimation thereof is limited by the number of measurements within each {solution interval}. As such, we have no guarantee that our estimate of the corrected visibility covariance matrix is accurate. This problem can be alleviated, for example {by estimating the covariance matrix for the antenna gains themselves}, and use {it} to build the visibility covariance matrix: this effectively improves conditioning (cf Sec. \ref{section.algorithm}). 

\pg
This paper is split into four main sections. In Section \ref{sec.formalism}, we derive the Cov-Cov relationship. With its newfound insights, we propose quality-based weighting schemes with which to improve radio interferometric images in Section \ref{sec.DynamicRange}. We follow in Section \ref{section.algorithm} by showing how to estimate, from real data, the covariance matrix from which the quality-based weights are derived. Our approach seems to give good results. Finally, we close the paper on a discussion of the applicability and limitations of the quality-based weighting scheme.

%% file: source/matrix-formalism.tex

\begin{table}[t!]
\begin{tabular}{c l}
\hline
&\\
Scalars & \\
\cline{1-1}
&\\
$n_{\mathrm{pix}}$ & Total number of pixels in image-plane \\ 
                   &  {\& number of cells in $uv$-grid}\\
$n_{\mathrm{ant}}$ & Total number of antennas in the array\\
$n_b$     & Number of visibilities\\
$b$       & Index for a single visibility.\\
& Equivalent to $\left(pq,t\nu\right)$\\
$\tnuIndeix$ & Equivalent to $\left(t,\nu\right)$\\
&\\
Vectors & \\
\cline{1-1}
&\\
$\muely$  & Residual image vector, size $n_{\mathrm{pix}}$\\
$\bm{\epsilon}$       & {Vector of $\epsilon$, size $n_{\mathrm{pix}}$}\\ 
$\bm{\widetilde{\gamma}}$ & Contains  gain products, size $n_b$. See Eq. \ref{eq.residuals.definition}\\
$\mathbf{1}$ & Vector containing 1 in every cell, size $n_b$\\
$\bm{\deltu}_{bb'}$ & Vector of coordinates in differential Fourier\\ & plane, of length $3$.\\
$\bm{l}_d$          & Vector of sky coordinates, of length $3$.\\
&\\
Matrices&\\
\cline{1-1}
&\\
$\matVt$       & Visibility seen by a baseline $pq$ at time and \\
& frequency $t,\nu$. Size $2\times 2$.\\
$\Kmat{p,t\nu}^d$& Fourier kernel for direction $d$, antenna $p$ and\\
                 &  one $\left(t,\nu\right)$ pair. Size $2 \times 2$.\\
$\matJ{p}$     & Jones matrix for antenna $p$ for one $\left(t,\nu\right)$ pair.\\
& Contains the gains  information. Size $2 \times 2$\\
$\Bmat{}$        & Sky brightness distribution matrix, of size $2 \times 2$.\\
$\bm{N}$       & Noise matrix, of size $2\times 2$. Contains a single \\
& realisation $n$ of the thermal noise in each cell.\\
$\muelF$       & Fourier transform matrix, of size $n_{\mathrm{pix}}\times n_{\mathrm{pix}}$.\\
$\muelS_b$     & Baseline selection matrix, which picks out 1\\
&  visibility out of the full set. Size $n_{\mathrm{pix}} \times n_b$\\
$\muelC_b$     & $n_{\mathrm{pix}} \times n_{\mathrm{pix}}$ convolution kernel that defines\\
               &  the PSF.\\
$\muelF_{bb'}$ & Convolution matrix mapping one $\deltu\deltv$ to $\deltl\deltm$.\\
& The set of all $\muelF_{bb'}$ determines the noise-PSF.\\
               & Size $n_{\mathrm{pix}} \times n_{\mathrm{pix}}$.\\
&\\
\hline
\end{tabular}
\caption{Table recapitulating the meaning and dimensions of vectors and matrices used {in Sec. \ref{sec.dudv}}. Only scalars which give matrix dimensions or indices are given here.}
\label{tab.variables}
\end{table}

\section{Building the Noise Map}\label{sec.formalism}

\pg
In this section, we derive our first fundamental result: the Cov-Cov relationship, Eq. \ref{eq.covcov.matrix}, which describes how the statistics of residual visibilities (and thus the antenna calibration solutions, henceforth ``gains") relate to the statistics of the image plane, i.e. of images made using the associated visibilities. The dimensions of the matrices (denoted by boldface capital letters) and vectors (denoted by boldface lowercase letters) used in this paper are given in Table \ref{tab.variables}, along with the scalar numbers used to denote specific dimensions. All other variables are scalars.

\subsection{The Cov-Cov Relationship in the $\deltu\deltv$ plane}\label{sec.dudv}

\pg
Let us begin by defining visibility gains. Using the Radio Interferometry Measurement Equation formalism for a sky consisting of a single point source (\cite{1996A&AS..117..137H}, \cite{2011A&A...527A.106S}, and companion papers), we can write the following relation between the sky and the signal as measured by a single baseline at time $t$ and frequency $\nu$:
\begin{align}\label{eq.ME.scalar}
\matVt &= \sum_d \Kmat{p,t\nu}^d \matJ{p}^d \Bmat{\nu}^d \left(\matJ{q}^d\right)^H \left(\Kmat{q,t\nu}^d\right)^H + \bm{N} 
\end{align}

\pg
All the quantities above are $2\times2$ matrices. Eq. \ref{eq.ME.scalar}, implies a linear relationship between the coherency matrix $\Bmat{\nu}^d$ and the visibilities recorded by a given baseline ($\matVt$), with the addition of a thermal noise matrix $\bm{N}$, which is also of shape $2 \times 2$ and contains different complex-valued realisations of the noise in each cell. Since electric fields are additive, the sky coherency matrix can be described as the sum of the contributions from individual sources in directions $d$ in the sky. We also assume that the sky does not vary over time, i.e. that $\Bmat{\nu}^d$ is not a function of time. The Jones matrices ($\matJ{...}^d$) contain the antenna gain information in matrix form, while $\Kmat{...,t\nu}^d$ is the Fourier kernel.  
Let us limit ourselves to the scalar case, which corresponds to assuming that emission is unpolarised. We assume that $\Bmat{}^d = s_d \I$, where $s$ is the flux of our single point source and $\I$ is the $2\times 2$ identity matrix. We also assume that $\bm{J}_{p,t\nu} = g_{p,t\nu} \I$, where $\gptnu$ is the complex-valued gains of antenna $p$ at time $t$ and frequency $\nu$. This means that we assume that the gains are direction-independent, and so $\bm{J}_{...,t\nu}^d$ becomes $\bm{J}_{...,t\nu}$. Similarly, $\Kmat{p,t\nu}^d = k_{p,t\nu}^d \I$, the Fourier kernel in the direction of the source, $d$. $\bm{N}$ has 1 realisation of $\epsilon$ in each cell, where:
\begin{alignat}{2}
\epsilon \sim \mathcal{N}\left(0,\sigma\right) + i \mathcal{N}\left(0,\sigma\right) \label{eq.noise.distribution}
\end{alignat}
where $\sigma$ is the variance of the thermal noise. Let us denote each $\left(t,\nu\right)$ pair by $\tnuIndeix$, and ignore the sky's frequency-dependence. The following scalar formulation is then equivalent to Eq. \ref{eq.ME.scalar}:

\begin{align} 
V_{pq}^{\tnuIndeix} &= \left(\sum_d s_d  k_{p,\tnuIndeix}^d\overline{k_{q,\tnuIndeix}^d}\right) g_p^{\tnuIndeix}\overline{g_{q}^{\tnuIndeix}}+\epsilon\\
k_{p,\tnuIndeix}^d  &= \exp\left(2\pi i \left(u_{p,\tnuIndeix} l_d + v_{p,\tnuIndeix} m_d + w_{p,\tnuIndeix} \left(n_p-1\right)\right)\right)\label{eq.fourier.kernel}
\end{align}

\pg
Calibration is the process of finding an accurate estimate of $\gpt$ for all antennas $p$, at all times $t$ and frequencies $\nu$. Since we are in a direction-independent regime, the quality of our calibration then determines the statistical properties of the residual visibilities (and the image-plane equivalent, the residual image). The residual visibilities associated with calibration solutions are defined as our measured visibilities minus the gain-corrupted model visibilities.
$\Hatgpt$ then denotes our calibration estimate for $\gpt$. We now begin to limit the generality of our framework by assuming that all sufficiently bright sources have been modeled and subtracted: unmodeled flux is then negligible. We can then write the residual visibilities as:
\begin{align}
\Rpqt =& \sum_{d} s_{d} \left( k_{p,\tnuIndeix}^{d} \overline{k_{q,\tnuIndeix}^{d}}\right) \left(\gpt \overline{\gqt} - \Hatgpt \overline{\Hatgqt}\right)  + \epsilon  \label{eq.residual.scalar}
\end{align}

The flux values in the image-plane pixels\footnote{{As opposed to the Fourier-plane pixels, which are the elements of the grid onto which the measured visibilities are mapped for imaging.}} are the Fourier transform of the visibility values mapped onto each pixel. This can be written as follows:
\begin{alignat}{2}
\displaystyle\muely &= \begin{pmatrix} \vdots \\ \sum_{pq} \Irpqlm \\ \vdots \end{pmatrix} \label{eq.def.muely}\\
\Irpqlm 
        &= \sum_\tnuIndeix \weight{\tnuIndeix} \Rpqt \kpqtlm \label{eq.residual.image}
\end{alignat}
where $lm$ are the directional cosine positions of a given pixel, and $\kpqtlm=k_{p,\tnuIndeix}^d\overline{k_{q,\tnuIndeix}^d}$ the Fourier coefficient mapping a point in Fourier space to a point on the image-plane. $\weight{\tnuIndeix}$ is the weight associated to a given visibility.
\pg
Let us now write this using a matrix formalism. The contribution of a single visibility $b=\left(pq,\tnuIndeix\right)$ to the image-plane residuals can be written as:
\begin{align}
\muely_b =& \muelF^H \muelS_b \omega_b \left(\kappa_b \matGains + \bm{\epsilon} \right)\\
\muely   =& \sum_b \muely_b\label{qwdqwdq}
\end{align}

%


where $\bm{\epsilon}$ is a vector of the $\epsilon$ of Eq. \ref{eq.noise.distribution} and $\muely$ is a vector of size $n_{\mathrm{pix}}$, with
\begin{align}	 
\kappa_{b} &= \sum_{d} s_{d} \kappa_{b}^{d}\label{eq.totalflux}\\
\kbd{b} &= k_{p,\tnuIndeix}^d \overline{k_{q,\tnuIndeix}^d} = \kpqtlm \label{eq.kpq}\\
\widetilde{\gamma}_b &= \gpt\overline{\gqt}-\Hatgpt\overline{\Hatgpt} \label{eq.residuals.definition}\\
\matGains &= \begin{pmatrix} \vdots \\ \widetilde{\gamma}_b \\ \vdots \end{pmatrix}
\end{align}
and $w_b$ is the scalar weight associated with each visibility. By default, $w_b = \frac{1}{n_b}$: all visibilities then have the same weight, and $\muely$ then becomes the average of all $\muely_b$. $\matGains$ is a {vector of all $\widetilde{\gamma}_b$, and thus of size $n_b$}. $\muelF$ is the Fourier kernel, of size $n_{\mathrm{pix}}\times n_{\mathrm{pix}}$. $\matS_b$ is a matrix of size $n_{pix}\times n_b$: its purpose is to encode the $uv$-coverage. Each $\matS_b$ contains only a single non-zero cell, different for different $\matS_b$. The height (number of rows) of $\matS_b$ is determined the size of the $uv$-grid, and its length (number of columns) by the number of visibilities. 

\pg
The order of operations is thus: each residual visibility $\left(\kappa_b \matGains + \bm{n} \right)$ is assigned some weight $ w_b$ and its $uv$-coordinates are set by $\muelS_b$. The inverse Fourier transform ($\muelF^H$) is then applied to this grid, and so we recover its image-plane fringe. By averaging over all fringes, we recover the dirty image.

%
%
%

\pg
The residual image will thus depend on three quantities: the residual gains, the flux in the image, and the weighting scheme. Let us consider the relationship between the statistics of residual visibilities and the variance at a given point in the corresponding residual image.

\subsection{Statistical Analysis}

\begin{figure*}[!t]
\centering
\includegraphics[width=\textwidth]{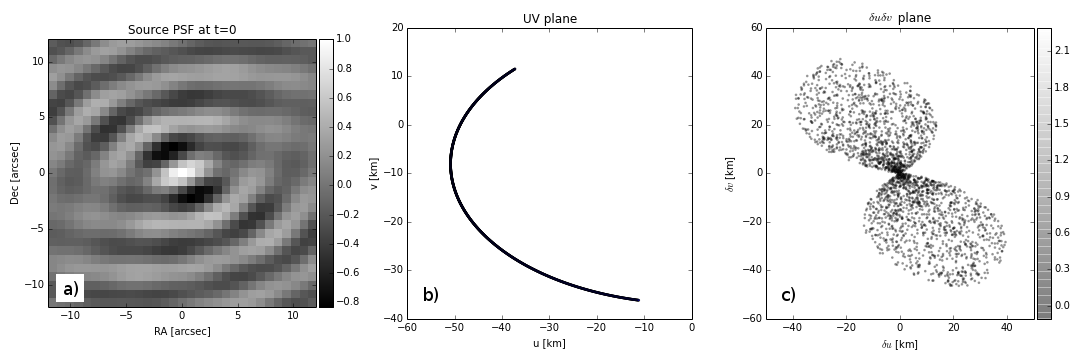}
\caption{\label{plot.psf.uv.dudv} Fig. \ref{plot.psf.uv.dudv}a shows the PSF image of a simulated 1Jy source at phase centre. {Colourbar units are in Jansky}.  Fig. \ref{plot.psf.uv.dudv}b shows the associated UV track, and Fig. \ref{plot.psf.uv.dudv}c the corresponding $\left(\deltu,\deltv\right)$ tracks. Note that the $\deltu\deltv$ plane does not have a homogeneous point density, but is denser near its origin: here, this is shown by plotting only 1 random point in 10000.}
\end{figure*}

\pg
In the following analysis, we treat our gain solutions and thermal noise as random variables in order to
compute the covariance matrix of our residual image, $\Cov{\muely}$. The diagonal of this matrix gives the variance for each pixel, while the wings give the covariance between pixels. Using the property that $\Cov{\bm{A}\bm{x}} = \bm{A}\Cov{\bm{x}}\bm{A}^H$, we can apply the $\Cov{}$ operator to Eq. \ref{qwdqwdq} to write:
\begin{align}
\Cov{\muely} =& \sum_{bb'} \muelF^H \underbrace{w_b w_{b'} \kappa_b \overline{\kappa_{b'}}}_{\bydefn \phi_{bb'}} \matS_b \Cov{\matGains} \matS_{b'}^T \muelF\\
              &+ \sum_b w_b^2 \sigma^2 \underbrace{\muelF^H \matS_b \I \matS_b^T \muelF}_{\bydefn\muelC_b}\\
             =&  \sum_{bb'} \phi_{bb'} \muelF^H\matS_b \Cov{\matGains} \matS_{b'}^T \muelF + \sum_b w_b^2 \sigma^2 \muelC_b
\end{align}
So far, we have only applied definitions. The net effect of $\matS_b \Cov{\matGains} \matS_{b'}^T$ (dimensions of $n_{\mathrm{pix}}\times n_{\mathrm{pix}}$) is to encode where a given baseline samples the $uv$-plane, and map {\emph{one} cell at matrix coordinates $(b,b')$} from the correlation matrix $\Cov{\matGains}$ onto the visibility grid. $\muelS_b$ is not the gridding kernel, but rather the sampling matrix, which determines where we have measurements and where we do not. We can thus write that $\matS_b \Cov{\matGains} \matS_{b'}^T=[\Cov{\matGains}]_{bb'} \matS_b \bm{1} \bm{1}^T \matS_{b'}^T$, where $[\Cov{\matGains}]_{bb'}$ is the value from the appropriate cell and $\bm{1}$ is the vector-of-ones of appropriate length (here, $n_b$). This allows us to write:
\begin{align}
\Cov{\muely} =& \sum_{bb'} \phi_{bb'} [\Cov{\matGains}]_{bb'} \muelF_{bb'} + \sum_b w_b^2 \sigma^2 \muelC_b\\
\mathrm{with}\qquad \muelF_{bb'} =& \left(\muelF_b\right)^H \underbrace{\muelF_{b'}}_{\mathclap{\bydefn \bm{1^T}\matS_{b'}^T\muelF}}
\end{align}
Here, $\muelC_b$ is a Toeplitz matrix, i.e. a convolution matrix, associated with baseline $b$. The set of all $\muelC_b$ defines the convolution kernel which characterises the Point-Spread Function (henceforth PSF) associated with a given uv-coverage, of size $n_{\mathrm{pix}}\times n_{\mathrm{pix}}$. $\muelF_{bb'}$, meanwhile, is not {generally} Toeplitz. Its cells can be written as:
\begin{align}
\muelF_{bb'}[d,d'] &= e^{2i\pi  \left(u_b l_d - u_{b'} l_{d'} + v_b m_d - v_{b'} m_{d'} + \left(n_d-1\right)w_b -\left(n_{d'}-1\right) w_{b'}\right)}
\end{align}
Let us investigate how the sky brightness distribution (i.e. $d$-dependence) affects the noise-map. We can write the sum over $bb'$ as two sums: one over $b,b'=b$ and one over $b,b'\ne b$. Thus:
\begin{align}
\Cov{\muely} =& \sum_b \left(\phi_{bb}  [\Cov{\matGains}]_{bb} + w_b^2 \sigma^2 \right) \muelC_b\nonumber\\
              &+ \sum_{b,b'\ne b} \phi_{bb'} [\Cov{\matGains}]_{bb'} \muelF_{bb'} \label{eq.cov.muely}
\end{align}
Note that the only direction-dependent terms in the above are $s_d$ and $\kappa_b^d$, which are both inside $\phi_{bb'}$ (for both $b=b'$ and $b \ne b'$). {By making the approximation that the Fourier kernels of different sources are orthogonal (i.e. that $\kappa_b^d \kappa_{b'}^{d'} = (\kappa_b^d)^2 \delta_{bb'}$)}\footnote{{This hypothesis is equivalent to assuming that the sky is dominated by distant point sources, where ``distant" means that the sources are multiple PSF Full-Width Half-Maximum apart}} we can write:
\begin{align}
\phi_{bb'} &= w_b w_{b'} \kappa_b \overline{\kappa_{b'}}\\
           &= w_b w_{b'} \left(\sum_d s_d \kappa_b^d\right)\left(\sum_{d'} s_{d'} \overline{\kappa_{b'}^{d'}}\right)\label{eq.dudv.intro}\\
           &\approx \sum_d w_b w_{b'} s_d^2 \kappa_b^d \overline{\kappa_{b'}^d}\label{whocares}\\
\phi_{bb'} &\approx \sum_d \phi_{bb'}^d
\end{align}
{Note that $\muelF_{bb}=\muelC_b$, since those are the coordinates along the diagonal: for these values, the matrix-of-ones at the centre of $\muelF_{bb'}$ becomes the identity matrix. Note also} that $\phi_{bb}^d = w_b^2 s_d^2$, since $\kappa_b^d \overline{\kappa_{b}^d}=1$. We can then write Eq. \ref{eq.cov.muely} as:
\begin{align}
\Cov{\muely} =& \sum_d \Bigg(\sum_b \phi_{bb}^d \left( [\Cov{\matGains}]_{bb} + \frac{w_b^2 \sigma^2}{\phi_{bb}} \right) \muelC_b\nonumber\\
              &+ \sum_{b,b'\ne b}  \phi_{bb'}^d [\Cov{\matGains}]_{bb'} \muelF_{bb'} \Bigg)\label{eq.covcov.matrix}
\end{align}
{where we have now limited our formalism to the case where the sky is dominated by distant point-like sources.}
\pg
This is our fundamental result: {assuming unpolarised emission coming from distant point sources and normally-distributed thermal noise}, it gives a direct relationship between the covariance of the residual visibilities and the covariance of the residual image-pixel values. We thus call it the Cov-Cov relationship. It describes the statistical properties of the image-plane as the result of a convolution process {changing an average noise level at different points in the image-plane, allowing} us to describe the behaviour of variance and covariance in the image. Let us focus on the first. 

\pg
By applying the $\diag{}$ operator (which returns the diagonal of an input matrix as a vector) to both sides of Eq.\ref{eq.covcov.matrix}, we can find an expression for the variance map in the image-plane:
\begin{align}
\Var{\muely}      =& \diag{\Cov{\muely}}\\
                  =& \sum_d \bigg(\sum_b  \left(\phi_{bb}^d  [\Cov{\matGains}]_{bb} + w_b^2 \sigma^2 \right) \underbrace{\diag{\muelC_b}}_{= \bm{1}} \nonumber\\
                   &\quad+  \sum_{b,b'\ne b}  \phi_{bb'}^d [\Cov{\matGains}]_{bb'} \diag{\muelF_{bb'}}\bigg)\label{eq.variance.imgplane}\\
\mathrm{where}\hspace{0.6cm}\bm{l}_d =& \left({l_d,m_d,\left(n_d-1\right)}\right)\\
\bm{\deltu}_{bb'} =& \left({\deltu}_{bb'},{\deltv}_{bb'},{\delta w}_{bb'}\right)\\
 				  =& \left( u_b - u_{b'},v_b-v_{b'},w_b-w_{b'}\right)
\end{align}
In Eq \ref{eq.variance.imgplane}, we have:
\begin{align}
\diag{\muelF_{bb'}}[d] &= e^{2i\pi \bm{l}_d \bm{\cdot} \bm{\deltu}_{bb'} }
\end{align}
{For $b\ne b'$, the diagonals of $\muelF_{bb'}$ are the Fourier kernels mapping $\deltu \deltv$ space to $\deltl\deltm$. $\muelF_{bb'}$ can then be thought of as a Fourier transform.} {It is not a diagonal matrix.}
It behaves as a \emph{covariance fringe}, allowing us to extend standard interferometric ideas to covariance space: each fringe can be thought of as a single ``spatial filter" applied to the pixel covariance matrix. Just as a given baseline has coordinates in $uv$-space, a given \emph{correlation between baseline {residual errors}} has coordinates in $uv$ \emph{correlation space}, which we will henceforth refer to as $\deltu\deltv$-space.

\pg
This $\deltu\deltv$ space warrants further discussion: Fig. \ref{plot.psf.uv.dudv} shows, for a given $uv$-track (Fig. \ref{plot.psf.uv.dudv}b), both the corresponding $\deltu\deltv$ domain (Fig. \ref{plot.psf.uv.dudv}c) and point-spread function (Fig. \ref{plot.psf.uv.dudv}a). {The symmetric, negative $uv$-track is treated as a separate track, and thus ignored. This means that we do not fully constrain the noise-PSF (since the covariance matrix of the symmetric track is simply the Hermitian of the first), but we do not seek to constrain it in this section, but rather to show that our results hold.}. We can see that the $\delta u \delta v$-tracks are symmetrical about the origin. The $\delta u \delta v$ space corresponding to a given $uv$-track can thus be most concisely described as a ``filled $uv$-track", with its boundaries defined by the ends of the $uv$-track. The set of $\muelF_{bb'}$, each of which maps one value of the covariance matrix to a fringe in the image-plane, would then characterise a PSF equivalent for the noise distribution, which we refer to as the noise-PSF. In our formalism, the only source of statistical effects in the field are calibration errors and thermal noise. The average variance in all pixels will be given by the diagonal of the covariance matrix and the thermal noise, provided that it truly follows a normal distribution. The only effects which will cause the variance in the image-plane to vary from one pixel to the next are those mapped onto the covariance fringes, i.e. such position-dependent variance fluctuations will be caused by correlated gain errors, which are spurious signal introduced by erroneous gain estimates. Assuming all sources in the field are point-like and distant, then these variance fluctuations will follow a specific distribution, convolved to every source in the field.
{Since the variance fluctuations act as tracers for calibration artefacts, artefacts in the image can be understood as \emph{one realisation} of the variance map, which is characterised by an average level determined by the variance in the gains and thermal noise, and a noise-PSF convolved with the sky brightness distribution}. The actual artefacts in the image will still be noisy, as a realisation of the true variance map. For the same reason, in the absence of correlated gain errors, $[\Cov{\matGains}]_{b \ne b'}$ are all zero and $\Cov{\matGains}$ is a diagonal matrix. {We then recover a ``flat" noise-map: the variance will be the same in all pixels, as the noise-PSF is absent}. In the ideal case, were we to recover the true value of the gains for all times and frequencies, this becomes pure thermal noise. 

%% file: source/matrix-noise-map-simulations.tex
\subsection{Noise Map Simulations}\label{sec.simulations}

\begin{figure}[t!]
\centering
\includegraphics[width=0.5\textwidth]{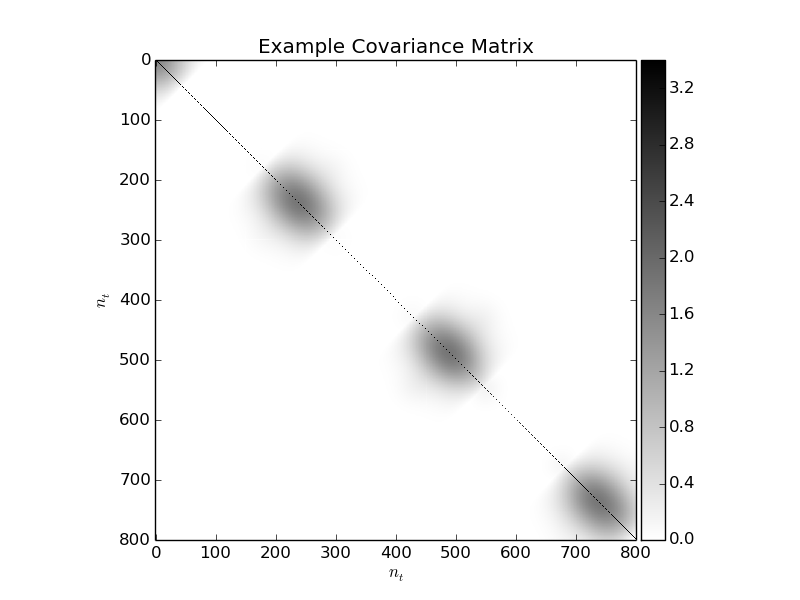}
\caption{\label{fig.covmat} Example of {a non-stationary covariance matrix}, which can be used to simulate $\Cov{\matGains}$.{The colourbar units are Jy$^2$}. The correlation scale $\sigma_\tnuIndeix$ is 40 cells, and the variability period is 500 cells. The matrix is made positive semi-definite (and therefore a covariance matrix) through SVD decomposition. The maximum size of the ``bubbles" is determined by $\sigma_\tnuIndeix$.}
\end{figure}

\pg
We have shown in Eq. \ref{eq.covcov.matrix} that there exists an analytical relationship between residual visibility statistics and image-plane residual statistics. This section gives details of simulations we have performed to support our claims on this ``Cov-Cov" relationship. Specifically, we simulate residual visibilities for a single baseline by generating a set of correlated random numbers with zero mean and a distribution following a specified covariance matrix $\bm{C}$. 
It contains a periodic function of period $T$ along the diagonal, which is then convolved with a Gaussian of width $\sigma_\tnuIndeix$ corresponding to the characteristic scale of correlation. The values of these parameters are chosen arbitrarily. A small constant term is added on the diagonal, the net value of which is strictly positive. This simulates a low thermal noise. Finally, singular value decomposition is used to ensure that this matrix is Hermitian positive semi-definite. The net effect is a {non-stationary} correlation: some residuals are correlated with their neighbours, and uncorrelated with others. An example of this covariance matrix for arbitrary parameter values is shown in Fig. \ref{fig.covmat}. We see that, for any given point, correlation is stronger with some neighbours than others (as determined by $\sigma_\tnuIndeix$).
Samples are drawn as follows: $\bm{C}$ is built following user specifications as described above. We then find its matrix square root $\bm{C}_0$ so as to apply it to random numbers generated from a normal distribution. We generate 2000 realisations $i$ of our random variables $\bm{r}_i$: 
\begin{alignat}{2}
\muely_i &= \sum_b \muelF_b \bm{r}_i \quad &\mathrm{with} \quad \bm{r}_i =& \bm{C}_0 \bm{x}\\
         &                                 &\mathrm{and} \quad \bm{x} \leftarrow& \mathcal{N}\left(0,1\right)\\
\muely   &= \begin{pmatrix} \hdots \muely_i \hdots \end{pmatrix}
\end{alignat}
{where $\muely$ is a matrix of dimensions $n_{\mathrm{realis}}\times n_b$. Since $\bm{x}$ follows a normal distribution, $\Cov{x}=\I$ and the covariance matrix of each $\muely_i$ is, by construction, $\bm{C}=\bm{C}_0 \bm{C}_0^H$. The covariance matrix of $\muely$ is thus also $\bm{C}$.}

\pg
As for the $uv$-track, our simulations read a single one from a specified dataset. In this case, we read an 8-hour $uv$-track for a baseline between two arbitrary LOFAR stations (specifically, CS001HBA0 and RS310HBA) in an observation of the Bootes extragalactic field. The effective baseline length varies between 37.9km and 51.8km. The dataset included 20 channels, each with a spectral width of 97.7 kHz; the central observing frequency is 139 MHz. The temporal resolution is 1 measurement per second.

We compare the \emph{measured} variance map $\bm{V}_y^m$, built by measuring the variance across realisations at each pixel in the image-plane, with the \emph{predicted} variance map $\Var{\muely}$, built using the Cov-Cov relationship (Eq. \ref{eq.covcov.matrix}). Since we are only interested in the variance map, rather than the covariance between pixels, we compute only the diagonal terms.
\begin{alignat}{2}
\bm{V}_{\tilde{y}}^m &= \diag{\muely\muely^H} \label{eq.meas.var}\\
\bm{V}_{\tilde{y}}^{pr} &= \sum_b [\bm{C}] _{bb}\I + \sum_{b,b'\ne b}\diag{\muelF_{bb'}}[\bm{C}]_{bb'}\label{eq.theo.var}
\end{alignat}
where the thermal noise is already incorporated into the diagonal of $\bm{C}$ and Eq. \ref{eq.theo.var} is merely the diagonal operator applied to the Cov-Cov relationship. 

\pg


\begin{figure*}[t!]
\centering
\includegraphics[width=\textwidth]{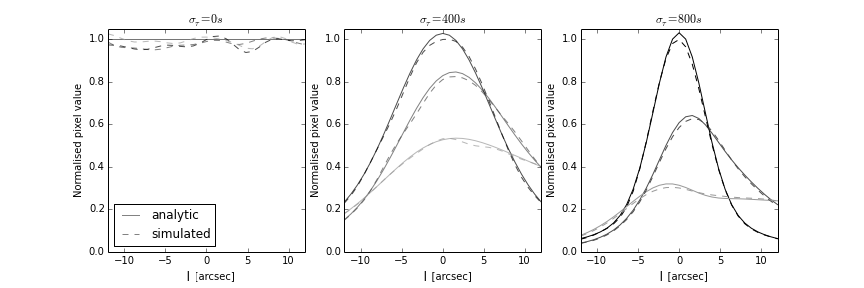}
\caption{\label{plot.cross-sections} The three lines in each figure correspond to three horizontal cross-sections from images in Fig. \ref{plot.noisepsfs}. {The units on the y-axis are dimensionless [$Jy^2/Jy^2$].} $\sigma_\tnuIndeix$ is the maximum characteristic error correlation length. In decreasing intensity, they correspond to  $m=0''$, $m=4''$, and $m=8''$. The dashed lines correspond to the variance measured with 2000 realisations for each pixel, while the solid line corresponds to the predicted value at that pixel. There are 31 pixels. We do not show cross-sections for negative $m$ due to image symmetry about the origin.}
\end{figure*}

\begin{figure}[t!]
\centering
\includegraphics[width=0.5\textwidth]{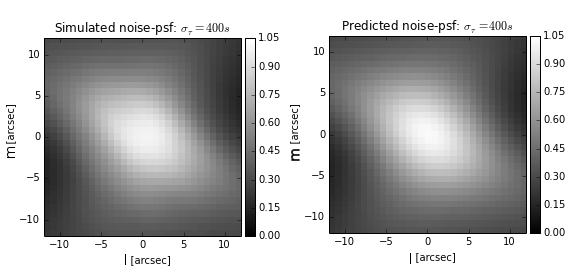} 
\caption{\label{plot.noisepsfs} Simulated noise-maps, compared with theoretical prediction. The pixel values are normalised by the average value of the covariance matrix: the units of the colourbar are thus dimensionless ($Jy^2/Jy^2$). These are on the same angular scale as the source shown in Fig. \ref{plot.psf.uv.dudv}.}
\end{figure}

\subsubsection{Simulation with a single point source}

\pg
We model our sky as containing a single 1 Jy point source at phase centre: we thus have $\phi_{bb}=w_b^2$. The source as seen through the set of $uv$-tracks used in our simulation, along with their corresponding $\left(\deltu,\deltv\right)$ space, are shown in Fig. \ref{plot.psf.uv.dudv}. The \emph{simulated} noise-map is calculated by drawing a large sample ($n_\mathrm{realis}=2000$) of random numbers from the correlated distribution, thereby creating 2000 sets of residual visibilities. By Fourier-transforming the visibilities to the image-plane and taking the variance of the values for each image pixel (i.e. each $l,m$ pair) as per Eq. \ref{eq.meas.var}, we can estimate $\Var{\muely}$. The \emph{predicted} noise-map, meanwhile, was found by assigning each cell of $\bm{C}$ to the appropriate point in the $\left(\deltu,\deltv\right)$ plane and Fourier transforming from this plane into the image-plane, as per Eq. \ref{eq.theo.var}.
We compare the outcome of simulating a large number $n_\mathrm{realis}$ of realisations and taking the variance across these realisations for each pixel with mapping the covariance matrix onto the $\deltu\deltv$-plane and using the Cov-Cov relationship.The results of our simulations are shown side-by-side in Fig. \ref{plot.noisepsfs}: the predicted and simulated noise-PSFs match. The peak-normalised predicted noise-PSF is less noisy, as shown in Fig. \ref{plot.cross-sections} for different correlation scales. This is expected, since it is calculated directly from the underlying distribution, rather than an estimate thereof. As $n_\mathrm{realis}\rightarrow\infty$, we expect the two methods to fully converge. As the maximum characteristic correlation length $\sigma_\tnuIndeix$ increases, the variance becomes ever more sharply peaked.

\pg
{Since our simulated sky consists of a 1Jy source at phase centre, there is only one noise-PSF to modulate the average noise-map, and it lies at phase centre. Let us test our formalism further by considering a model with multiple point sources.}

\subsubsection{Simulation with 3 point sources}

\begin{figure}[t!]
\centering
\includegraphics[width=0.5\textwidth]{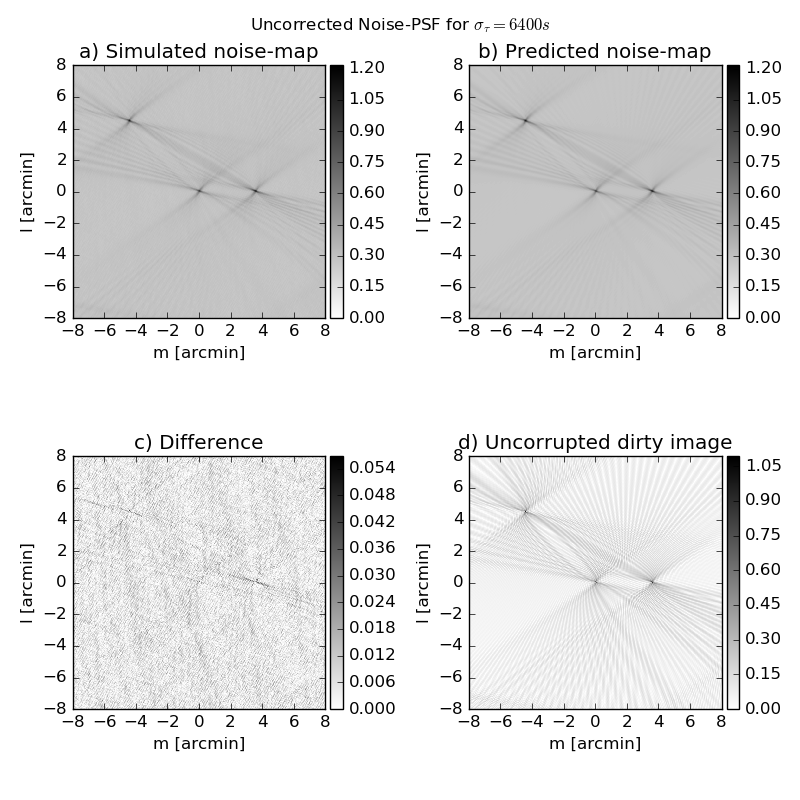}
\caption{\label{imag.simu-3sources.noisemap} {Noise-map of sky with correlated gain errors and three point sources. The colourbars of (a), (b) and (c) have dimensionless units, while that of (d) is in Jansky. Note the presence of structure in the residuals (c): these show the limits of our hypothesis that sources are spatially incoherent.}}
\end{figure}

\pg
{We wish to test our prediction that the noise-map can be described as a convolutional process modulating an average noise level. We thus perform another simulation, this time with three 1Jy point sources. The associated dirty image is shown in Fig. \ref{imag.simu-3sources.noisemap}d. }

\pg
{This dirty image simply consists of performing a direct Fourier transform (i.e. without using a Fast Fourier Transform algorithm) on simulated visibilities corresponding to these three point sources. We now perform a similar test as above on this field. Firstly, we ``apply" gain errors to these visibilities by multiplying our model with our residual gain errors. This allows us to find 2000 realisations of residual visibilities, and find the variance for each pixel across these realisations. This gives us the simulated noise-map, shown in Fig. \ref{imag.simu-3sources.noisemap}a. Secondly, we perform a DFT from the differential Fourier plane to the $(l,m)$ plane as before, assigning one cell of $\Cov{\matGains}$ to each point of the differential Fourier plane. This time, however, $\phi_{bb'}^d$ is not simply unity for all points in the differential Fourier plane. Instead, it is calculated for the three-point-source model, and applied for each point. This gives us the predicted noise-map in Fig. \ref{imag.simu-3sources.noisemap}b. Finally, Fig. \ref{imag.simu-3sources.noisemap}c shows the absolute value of the difference between the two images. We see that there is some structure present in these residuals: this is expected, as the PSF of the sources in the dirty imags clearly overlap. We are thus not quite in the regime where emission is fully spatially incoherent. Nevertheless, our predictions hold to better than $5\%$ accuracy. }

\pg
It bears repeating that, for correlated noise, this map can be understood as a distribution map for calibration artefacts: the amount of spurious correlated emission seen by each baseline will determine the noise-map, and the true image-plane artefacts will then be \emph{one set of realisations} of this \emph{underlying distribution}.

%% file: source/flattening.tex
\section{Adaptive Quality-based Weighting Schemes}\label{sec.DynamicRange}

\pg
As discussed in Section \ref{sec.intro}, some intervals of an observation will have lower gain variability. These will show up in the gain covariance matrix as intervals with lower variance. Similarly, those with larger intrinsic gain variability will have greater error in their gain estimate. By giving greater weights to the former, and lower weights to the latter, we expect to be able to improve image reconstruction. {We thus talk of adaptive quality-based weighting, as the weights will adapt based on the calibration quality.}

\pg
The pixel variance is determined by the visibility covariance matrix, as shown in Eq. \ref{eq.variance.imgplane}. The diagonal of the visibility covariance matrix will add a flat noise to all pixels, while its wings will determine the calibration artefact distribution, which will be convolved to the sky brightness distribution. We thus have two sources of variance in the image-plane.
Minimising the far-field noise (i.e. the variance far from sources) in an image would involve down-weighting noisier calibration intervals while up-weighting the more quiescent ones, without taking noise-correlation between visibilities into account. This is because the far-field noise will be dominated by the diagonal component of the covariance matrix (cf. Eq. \ref{eq.variance.imgplane}). By the same token, minimising calibration artefacts would involve down-weighting measurements with strongly-correlated noise, and up-weighting the less-correlated. This would not, however, minimise the diagonal component: in fact, it will likely exaggerate its up-weighting and down-weighting. As such, it will increase the constant level of the noise-map, but flatten the noise-PSF's contribution. There are thus two competing types of noise that we seek to minimise: uncorrelated noise, which corresponds to $\dudv=0$ (i.e. the diagonal components of the gain covariance matrix), and correlated noise, which corresponds to $\dudv\ne 0$ (i.e. its wings). Minimising the first will minimise far-field noise without optimally reducing artefacts, while minimising the last will minimise noise near sources at a cost to far-field noise. In the following sub-sections, we will discuss weighting schemes used to accomplish this.

\subsection{Optimising sensitivity}\label{sec.lightweights.formalism}

\pg
The Cov-Cov relationship (Eq. \ref{eq.covcov.matrix}) tells us that, far from any sources, the variance map (Eq. \ref{eq.variance.imgplane}) is dominated by a constant term: the contribution from thermal noise and the diagonal of the residual visibility covariance matrix. Maximising sensitivity far from sources therefore implies minimising $\diag{\Cov{\matGains}}$. This is equivalent to assigning visibilities weights inversely proportional to their variance:
\begin{align}
w_{b} &= \frac{1}{\Var{\matGains_{b}}}
\end{align}

\pg
For each baseline, those times with larger variance in the residuals will be down-weighted, and those with smaller variance will be up-weighted; this scheme does not require information about the underlying gains, only the error on our solutions. Since we are treating $\sigma_n^2$ as a constant for all antennas and all times, those times where our gains estimate is closer to the true gains will be up-weighted, and those moments where they are farther from the actual gains will be down-weighted: hence the term ``adaptive quality-based weighting". Note that the diagonal of the weighted residuals' covariance matrix should therefore become constant: this weighting scheme explicitly brings the residuals closer to what is expected in the case of perfect calibration, assuming uncorrelated noise. For the remainder of this paper, we will refer to these weights as \emph{sensitivity-optimal} weighting.

\subsection{Minimising Calibration Artefacts}\label{sec.fullweights.formalism}

\pg
Minimising calibration artefacts - i.e. optimising the sensitivity near bright sources - means flattening the noise-map. Since the noise-map can be understood as a noise-PSF convolved with all the modeled sources in the sky {modulating the background variance level,} it will be flattest when its peak is minimised. From the Cov-Cov relationship (Eq. \ref{eq.covcov.matrix}), we can see that, at the peak of the noise-PSF (which would be the variance at the pixel where $l=m=0$), the Fourier kernel is unity: the variance for that pixel is thus the sum of all the cells in the covariance matrix. By accounting for normalisation, we can write the variance at the centre of the noise-PSF as:
\begin{align}
V\left(\bm{w}\right) = \frac{\bm{w}^T {\Cov{\matGains}}\bm{w} }{\bm{w}^T \mathbf{1} \mathbf{1}^T \bm{w}}
\end{align}
Our optimality condition is then, after some algebra:
\begin{alignat}{2}
0 &=\frac{\partial}{\partial \bm{w}}\left( V \right)\\
\leftrightarrow
\Cov{\matGains} \bm{w} &= \mathbf{1}\mathbf{1}^T \bm{w} \left(\bm{w}^T \bm{1}\bm{1}^T\bm{w}\right)^{-1} \bm{w}^T\Cov{\matGains} \bm{w}
\end{alignat}

\pg
We find that one $\bm{w}$ which satisfies the above is:
\begin{align}
\bm{w} &= \Cov{\matGains}^{-1} \mathbf{1}\label{eq.fullweights.analytical}
\end{align}
where $\mathbf{1}$ is a vector of ones. These weights depend only on calibration quality: badly-calibrated cells will include spurious time-correlated signal introduced by trying to fit the noise $n$ on visibilities. Down-weighting these cells helps suppress artefacts in the field, at the cost of far-field sensitivity. This weighting scheme is thus only a function of the relative quality of calibration at different times, boosting better-calibrated visibilities and suppressing poorly-calibrated visibilities. For the remainder of this paper, we will refer to these weights as \emph{artefact-optimal} weighting.

%% file: source/matrix-algorithm.tex
\section{Estimating the Covariance Matrix}\label{section.algorithm}
\pg
In our simulations, we have worked from a known covariance matrix and shown that our predictions for the residual image's behaviour hold. With real data, however, we do not have access to this underlying covariance matrix. Since our weights are extracted from said matrix, estimating it as accurately as possible remains a challenge: this is in turn limited by the number of samples which can be used for each cell.

\pg
Each cell in the covariance matrix is built by averaging a number of measurements, or samples. The more samples are available, the better our estimate becomes: once we have more samples than degrees of freedom, we say that our estimation is well-conditioned. Otherwise, it is poorly-conditioned. In this section, we will discuss ways in which we can improve the conditioning of the covariance matrix estimation.

\subsection{Baseline-based Estimation}

\pg
One way to improve the conditioning of our covariance matrix estimation is to make the same hypothesis as the calibration algorithm: we can treat the underlying gains as constant within each calibration interval. Provided this interval is known, this allows us to find a single estimate for each interval block of the covariance matrix, turning a $n_b\times n_b$ matrix into a smaller $n_{\mathrm{intervals}}\times n_{\mathrm{intervals}}$ equivalent, where $n_{\mathrm{intervals}}$ is the number of solution intervals used for to find the gain solutions. We then improve our conditioning by a factor of $n_\mathrm{int}$, which is the number of samples in a calibration interval. The estimate $\widehat{\Cov{\matGains}}$of the covariance matrix $\Cov{\matGains}$ is built by applying the covariance operator:
\begin{align}
\widehat{\Cov{\matGains}} \bydefn \estC =&  \frac{1}{n_\mathrm{int}}\sum_{i\in n_{\mathrm{int}}}\left(\matGains_i- \langle \matGains \rangle\right) \left(\matGains_i - \langle \matGains \rangle \right)^H
\end{align}
where the $\langle \hdots\rangle$ operator denotes taking the average over the full vector. If the calibration solver's gain estimates are unbiased (i.e. $\Exp{\ghatvect}=\gvect$) \emph{and} the model of the sky is sufficiently complete, this quantity should be zero. Having created $\estC$, which will be of size $n_b \times n_b$, its cells can now be averaged over blocks of $n_\mathrm{int}\times n_\mathrm{int}$. This allows us to estimate the weights for each baseline and each time.

\pg
Mathematically, we retrace the steps of Section \ref{sec.formalism}. In the absence of direction-dependent effects, we define the residual visibilities as before, and use them to define the normalised residual visibilities $\rho_b$:
\begin{align}
r_{b} =& w_b \kappa_b \tilde{\gamma}_b  + \epsilon \label{eq.residual.matrix}\\
\rho_{b} =& \frac{r_{b}}{k_b}
\end{align}
We then organise the residuals in cells:
\begin{align}
\Rmat{\cellindex} =&   \begin{pmatrix} \vdots \\ \rho_{b\in \cellindex} \\ \vdots \end{pmatrix}\\ 
\bm{R} =&   \begin{pmatrix} \hdots \quad \Rmat{\cellindex} \quad \hdots \end{pmatrix}
\end{align}

\pg
$\bm{R}$ corresponds to a matrix containing all the residual visibilities within one calibration cell $\cellindex$, i.e. for $b \in \cellindex$ where $\gvect_{\cellindex} = \const$ {It is therefore of size $n_{\mathrm{intervals}}\times n_{\cellindex}$, where $n_{\mathrm{intervals}}$ is the number of calibration intervals in the observation}. Normalising the residual visibilities by $k_b$ allows us to recover the underlying covariance matrix by multiplying the residual visibility matrix $\bm{R}$ with its Hermitian conjugate:
\begin{align}
\estC[b\in\cellindex,b'\in\cellindex']             =& \left(\bm{R}^H \bm{R}\right)[\cellindex,\cellindex']
\end{align}
Note that we have divided the noise term by the flux model $S_b$, which can be very small in some cells. As such, care must be taken not to cause the relative thermal noise contribution to explode: those cells where this would occur are dominated by thermal noise, and information on the covariance matrix cannot be recovered from them.

\pg
In this framework, we simply treat the index $\cellindex$ as containing all the times and frequencies, for individual baselines, corresponding to a single calibration interval. $\estC$ is then an estimate of the residual visibility covariance matrix.

\subsection{Antenna-based Estimation}

In the subsection above, we assume{d} that finding one solution per interval will give us strong enough constraints to make the problem {of estimating the covariance matrix} well-conditioned: this may not be true in all cases. Conditioning may then need to be improved further: in this subsection, we show one way in which this can be done. There are others, e.g. using the rank of the matrix itself to find better-conditioned estimates of the covariance matrix at a lower resolution (i.e. a single estimate for a greater number of cells). They will not be presented in this paper, but are a possible avenue future work on this topic.

\pg
In estimating the covariance matrix for each baseline and each calibration cell, we are severely limited by the small number of samples in each cell. One way to overcome this problem is to find estimates for the variance of \emph{antenna} gains, and use these to return to the baseline variances. In this formalism, we extend  $\cellindex$ to include all visibilities pointing at a single antenna at a given time. Let us begin by writing an expression for the gain vector, which contains the gains for all antennas and all calibration cells:
\begin{align}
\ghatvect_c =& \begin{pmatrix} \vdots \\ \hat{g}_p^{\tnuIndeix\in c} \\ \vdots \end{pmatrix}\\
\Ghatvect =& \begin{pmatrix} \hdots \quad \ghatvect_c \quad \hdots \end{pmatrix}
\end{align}
and the variance on each antenna in each calibration cell is then:
\begin{align}
\Var{\ghatvect_c} 
               =& \Exp{\ghatvect_c \ghatvect_c^H} - \underbrace{\Exp{\ghatvect_c} \Exp{\ghatvect_c}^H}_{= \gvect_c \gvect_c^H} \label{eq.def.vector.variance}
\end{align}

\pg
As we can see, Eq. \ref{eq.def.vector.variance} is simply a vector form of Eq. \ref{eq.residuals.definition}. 
The residual gains of Eq. \ref{eq.residuals.definition} can now be understood as random samples of the covariance between the gains for antennas $p$ and $q$ at a given time, assuming complete skymodel subtraction. We can thus define the variance sample matrix as an \emph{estimate} of the \emph{variance matrix}:
\begin{align}
\VarmatC =&\widehat{\Var{\gvect_c}} \\
                   =& \sum_{\tnuIndeix\in\cellindex} \left( \ghatvect_\tnuIndeix\ghatvect_\tnuIndeix^H - \gvect_\tnuIndeix\gvect_\tnuIndeix^H \right)
\end{align}

\pg
We define the residual matrix as:
\begin{align}
\Rmat{\tnuIndeix} =& \sum_d s_d \Kmat{d,\tnuIndeix} \left( \ghatvect_\tnuIndeix\ghatvect_\tnuIndeix^H - \gvect_\tnuIndeix\gvect_\tnuIndeix^H \right) \Kmat{d,\tnuIndeix}^H + \epsilon \label{eq.residual.matrix}
\end{align}
where we {explicitly place ourselves in the limits of our formalism, i.e.} that we do not have direction-dependent gains. We now see that at the core of Eq. \ref{eq.residual.matrix} lies $\displaystyle\VarmatTau$, where $\sum_{\tnuIndeix\in\cellindex} \VarmatTau = \VarmatC$.
The K-matrix is defined as follows:
\begin{align}\label{kmat.def}
\Kmat{d,\tnuIndeix} =&
\begin{pmatrix}
k_{p,\tnuIndeix}^{d}   &                       &   0     &\\
                       & k_{q,\tnuIndeix}^{d}  &        &\\ 
         0              &                       & \ddots 
\end{pmatrix}
\end{align}
Since the residual matrix depends on the gains, we define the residual visibility vectors as:
\begin{align}
\Rmat{{\cellindex}} =&   \begin{pmatrix} \vdots \\ \Rmat{\tnuIndeix\in \cellindex} \\ \vdots \end{pmatrix} \\
\bm{R} =&   \begin{pmatrix} \hdots \quad \Rmat{c} \quad \hdots \end{pmatrix}
\end{align}

\pg
$\Rmat{\cellindex}$ corresponds to a matrix containing all the residual visibilities within one calibration cell $\cellindex$, i.e. for $\tnuIndeix \in \cellindex$ where $\gvect_{\cellindex} = \const$ Let us define $n_\cellindex$ as the number of elements in each calibration cell. The residual variance sample matrix can now be built by multiplying the residual visibility matrix with its Hermitian conjugate:
\begin{align}
\matVV             =& \bm{R}^H \bm{R} \label{eq.matvv.definition} 
\end{align}
Note that we do this because it allows us to turn a single noise realisation $\epsilon$ into a statistical quantity $\sigma$.
We can relate $\matVV$ to the variance of individual antenna gains:
\begin{align}
\matVV  
            =& \sum_{\tnuIndeix\in\cellindex} \left(\sum_{d,d'} s_d \Kmat{d,\tnuIndeix} \left( \VarmatC \right)^H \Kmat{d,\tnuIndeix}^H s_{d'} \Kmat{d',\tnuIndeix} \left( \VarmatC \right) \Kmat{d',\tnuIndeix}^H + \I\sigma^2 \right)
\end{align}
To reach this point, in Eq. \ref{whocares}, we made the hypothesis that the sky brightness distribution is dominated by spatially incoherent emission. Applying this hypothesis again here, we can make the approximation that the cross-terms in the sum over $d,d'$ average to zero: $\sum_{d,d'\ne d}\approx 0$. We then have:
\begin{align}
\matVV 
      \approx& \sum_\tnuIndeix \left(\sum_d s_d^2 \Kmat{d,\tnuIndeix} \left( \VarmatC \right)^H  \left( \VarmatC \right) \Kmat{d,\tnuIndeix}^H + \I\sigma^2\right)\\
            =& \left(\VarmatC\right)^2\hadam \left(\underbrace{\sum_{\tnuIndeix}\sum_d s_d^2 \kvect{d,\tnuIndeix}\kvect{d,\tnuIndeix}^H}_{\bydefn \mathcal{S}}\right) + n_\cellindex\I\sigma^2\\
\VarmatC    =& \sqrt{\mathcal{S}^{\hadam-1}\left(\matVV-n_\cellindex\I\sigma^2\right)}
\end{align}
where $\hadam$ denotes the Hadamard or entrywise product and $\kvect{}=\diag{\Kmat{}}$. 
Thus, $\matVV$ allows us to estimate the variance of each antenna and for each calibration cell by using all the visibilities pointing to that antenna within that calibration cell. With this information, we can then rebuild the baseline-dependent matrix, having improved our sampling by a factor of $n_{\mathrm{ant}}$.

\begin{figure*}[t!]
\centering
\begin{subfigure}{.43\textwidth}
\resizebox{\hsize}{!}{\includegraphics{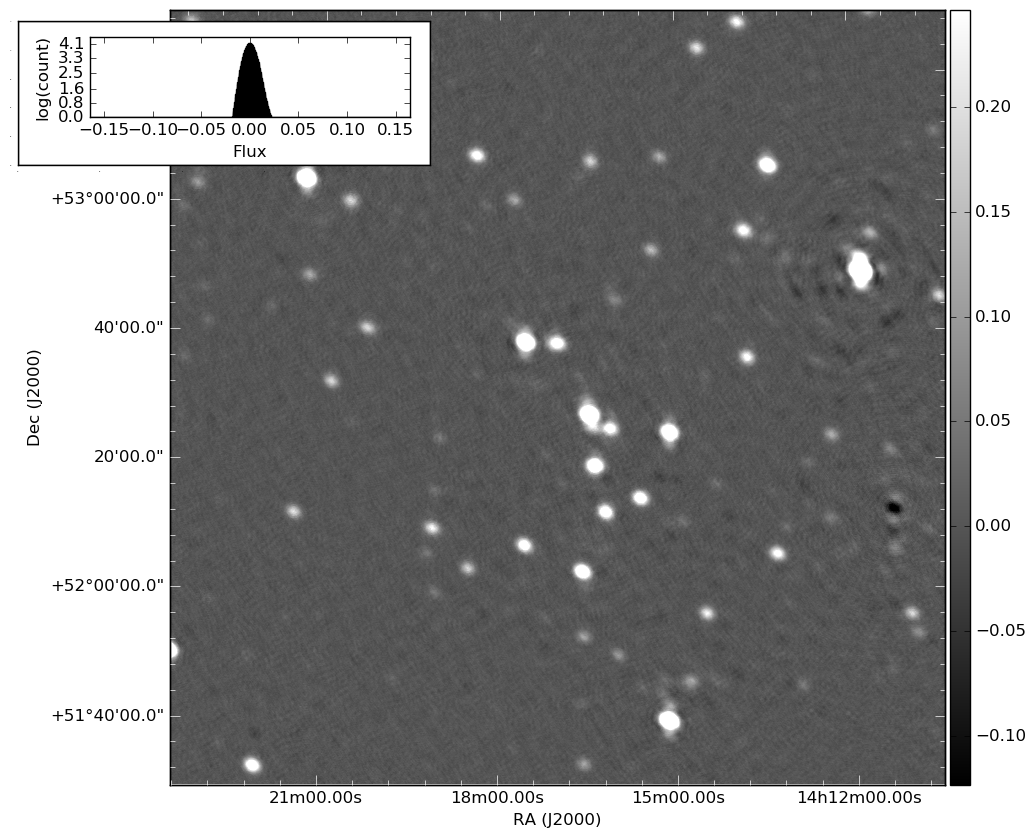}}
\caption{\label{image.3c295.goodcal} Well-calibrated, unweighted restored image of the sky near the centre of the Extended Groth Strip. Used for comparison with the other images. {Units of colourbar are Janskys}. This image was made with data calibrated following best practice (solution intervals of 8 seconds, half the bandwidth). rms=5.87mJy/beam}
\end{subfigure}
\hfill
\begin{subfigure}{.43\textwidth}
\resizebox{\hsize}{!}{\includegraphics{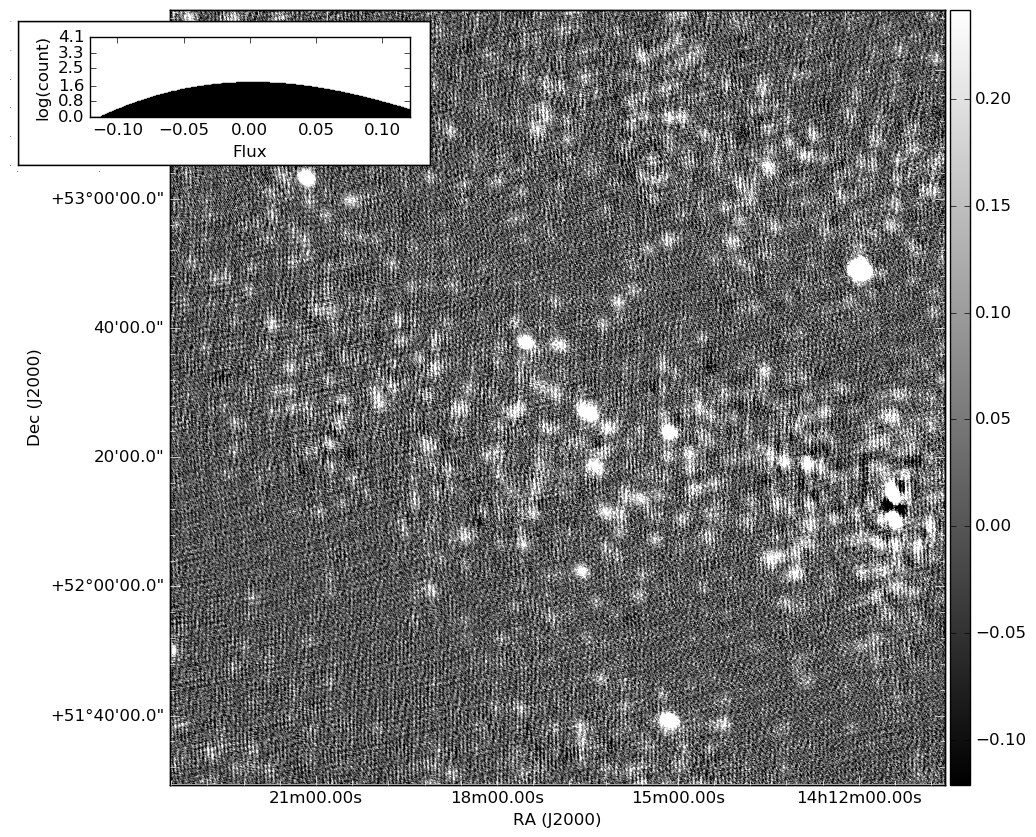}}
\caption{\label{image.3c295.nocorr} Poorly-calibrated, unweighted restored image of the sky near the centre of the Extended Groth Strip. {Units of colourbar are Janskys}. This image was made with the same data as for Fig. \ref{image.3c295.goodcal}, but averaged in time and calibrated using larger gain solution intervals: 2 minutes and half the bandwidth. rms=86.4mJy/beam}
\end{subfigure}
\hfill
\begin{subfigure}{.43\textwidth}
\resizebox{\hsize}{!}{\includegraphics{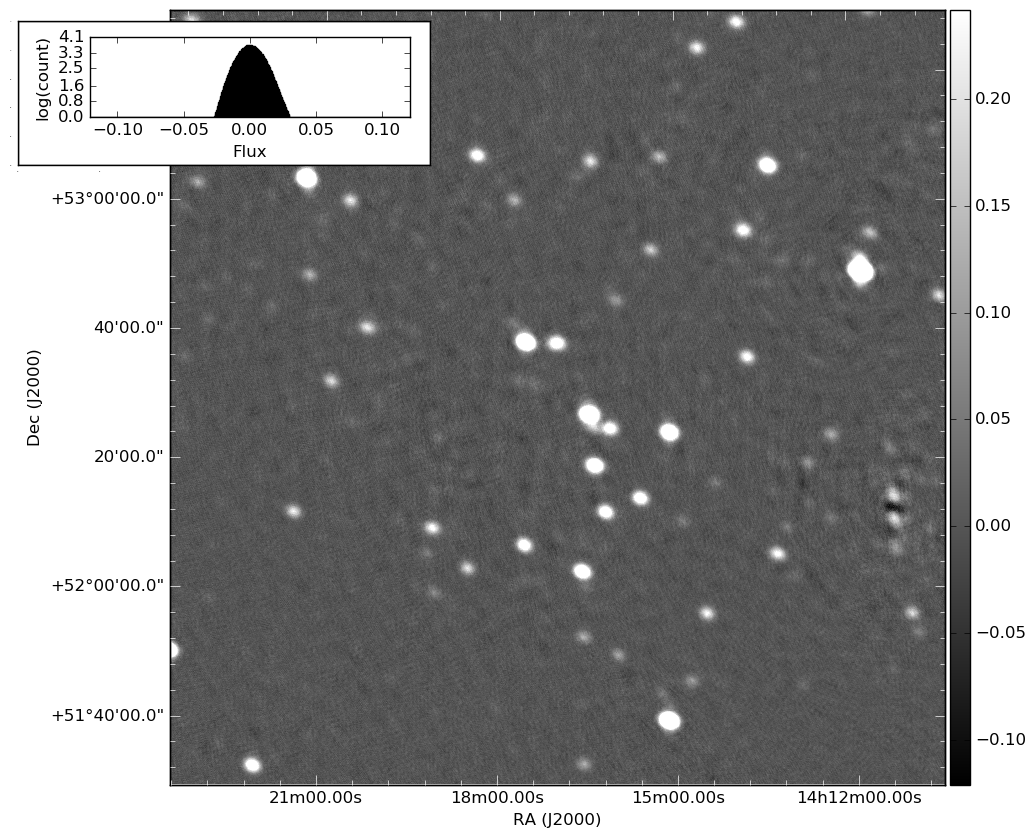}}
\caption{\label{image.3c295.lightcorr} Image made using the same imaging parameters and corrected visibilities as Fig. \ref{image.3c295.nocorr}, using sensitivity-optimal weighting. {Units of colourbar are Janskys}. rms=9.69mJy/beam}
\end{subfigure}
\hfill
\begin{subfigure}{.43\textwidth}
\resizebox{\hsize}{!}{\includegraphics{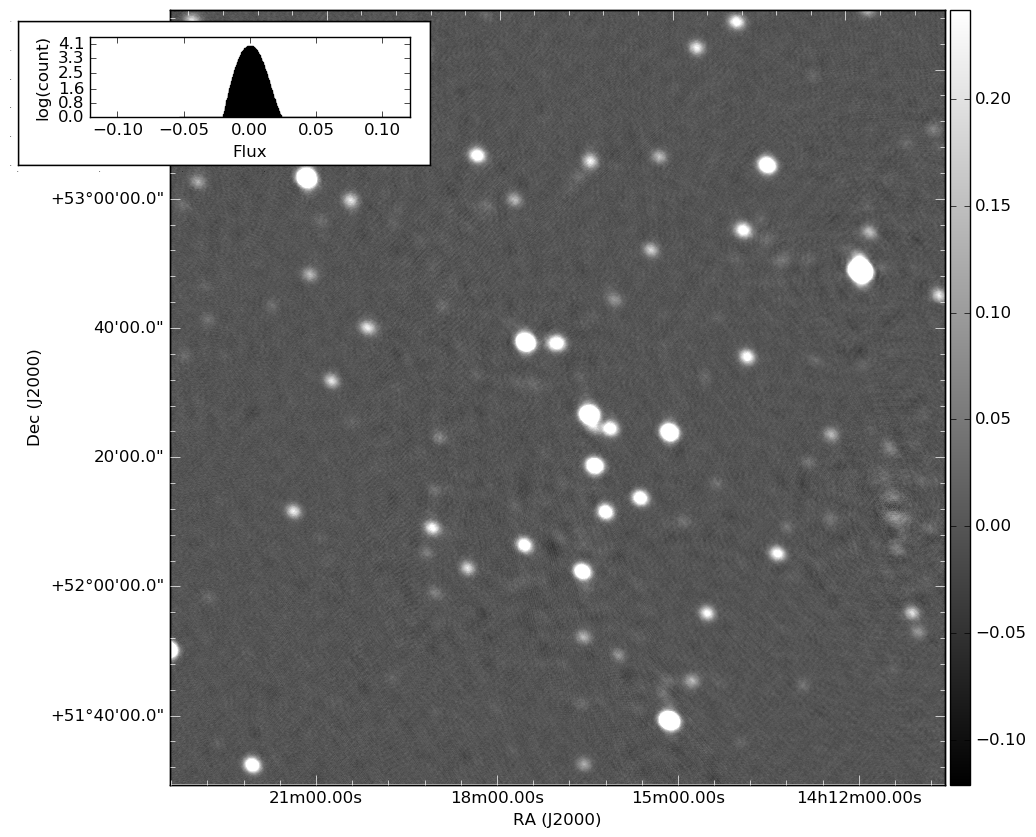}}
\caption{\label{image.3c295.fullcorr} Image made using the same imaging parameters and corrected visibilities as Fig. \ref{image.3c295.nocorr}, using artefact-optimal weighting. {Units of colourbar are Janskys}. rms=15.8mJy/beam}
\end{subfigure}
\caption{\label{image.fourRealImages}Restored images of the centre of the Extended Groth Strip, as seen with an 8-hour observation using the full LOFAR array. Fig. \ref{image.3c295.goodcal} shows an image of the field made with good calibration intervals. Fig. \ref{image.3c295.nocorr} shows an image of the field made with poor calibration intervals. Fig. \ref{image.3c295.lightcorr} shows image made with the same visibilities and imaging parameters, but with the application of the sensitivity-optimal weighting scheme. Fig. \ref{image.3c295.fullcorr}, similarly, differs from Fig \ref{image.3c295.lightcorr} only in that artefact-optimal weights, rather than sensitivity-optimal weights, were used. {The histograms of pixel values in each image have 1000 flux bins ranging from -0.16 Jy to 0.16 Jy. Their ordinates are in log scale.} Pixel size is 1.5'' in all images.}

\end{figure*}

\section{Applying the Correction to Simulated Data}\label{section.simulations.application}

\begin{figure}[!t]
    \centering
        \centering
        \includegraphics[width=\linewidth]{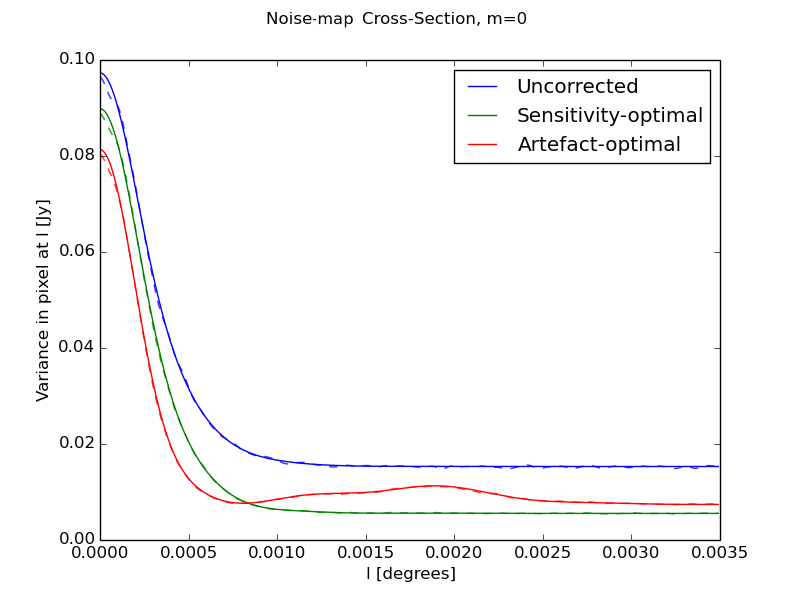} 
        \caption{The sensitivity-optimal (green) and artefact-optimal (red) weights both give improvements over the unweighted noise-map (blue).} \label{fig.simu.weightapplied}
\end{figure}

\pg
In this section, we show the impact of our weighting schemes on a noise-map made from arbitrarily strongly-correlated residuals. Here, we assume that our sky contains only a single point source at phase centre: there is thus only a single instance of the noise-PSF, placed at phase centre, to modulate the average variance level. We sample this instance by taking a cross-section from $\left(l,m\right)=0$ to a large $l$, keeping $m$ constant. The only difference between these cross-sections is the weighting scheme applied: unit weights for all visibilities (``Uncorrected", blue), sensitivity-optimal weights (green), and artefact-optimal weights (red). We plot both the result predicted by the Cov-Cov relationship (solid line) and the variance estimated across 2000 realisations (dashed line): the result is shown in Fig. \ref{fig.simu.weightapplied}. The two remain in such agreement throughout the cross-section as to be nearly indistinguishable. 

\pg
There are a few significant points to note on this figure. Firstly, most of the power in the matrix lies along the diagonal: both weighting schemes thus give good improvements in variance across the map. The artefact-optimal weights, while decreasing the peak further, as expected, also increases the noise far from sources: this is due to the fact that the artefact-optimal weights are in a sense more ``selective" than the sensitivity-optimal weights: they up-weigh and down-weigh more severely, and will only result in a constant covariance matrix if this matrix is zero everywhere outside of the diagonal. {In effect, the noise-map becomes flatter, but much broader.}

\section{Applying the Correction to Real Data}

\begin{figure*}
\begin{subfigure}{.43\textwidth}
\resizebox{\hsize}{!}{\includegraphics{images/paperfig-goodcal.png}}
\caption{\label{image.3c295.goodcal-2} Well-calibrated, unweighted restored image of the sky near the centre of the Extended Groth Strip. Used for comparison with the other images. Calibration solution intervals used were 8 seconds, half the bandwidth. rms=5.87mJy/beam}
\end{subfigure}
\hfill
\begin{subfigure}{.43\textwidth}
\resizebox{\hsize}{!}{\includegraphics{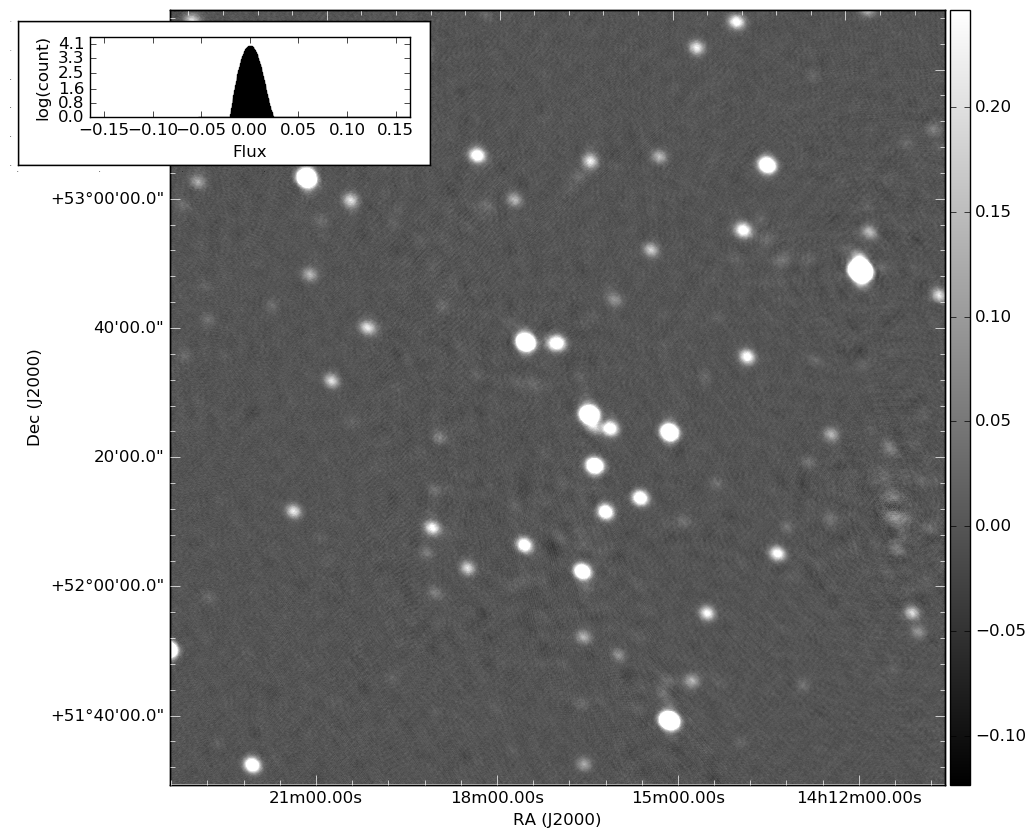}}
\caption{\label{image.3c295.antcorr} Image made using the same imaging parameters and corrected visibilities as Fig. \ref{image.3c295.nocorr}, with the application of antenna-based, sensitivity-optimal weighting. Solution interval of 2 minutes, half the bandwidth. rms=6.69mJy/beam}
\end{subfigure}
\caption{\label{image.twoRealImages}Comparison between the well-calibrated image (i.e. the same image as Fig \ref*{image.3c295.goodcal}) and antenna-based sensitivity-optimal weights.  {Units of both colourbars are Jansky}. We see that we recover a very similar image, despite the fact that the data used for the weighted image are averaged by a factor of 8 compared to those used for the unweighted image.}
\end{figure*}

\pg
In this section, we show the effect of adaptive quality-based weighting on real data. The dataset used in this section is a single sub-band from an 8-hour LOFAR observation centred on the Extended Groth Strip ($\alpha$=14:19:17.84,$\delta$=52:49:26.49). The observation was performed on August 28th, 2014. The subband includes 8 channels of width 24.414 kHz each, for a total bandwidth ranging from 150.2 to 150.5 MHz. The data have been averaged in time to 1 data point per second. The data was calibrated using Wirtinger calibration \citep[see][and references therein]{2014arXiv1410.8706T,2015MNRAS.449.2668S} and a sky model consisting only of a nearby calibrator source, 3C295. A reference image (a cutout of which is shown in Fig. \ref{image.3c295.goodcal}) was made by calibrating the data according to best practice for LOFAR survey data: 1 calibration solution per 8 seconds and per 4 channels. {The residual data was then corrected by the gain solutions and imaged} using Briggs weighting (robust=0), pixel size of $1.5''$, and deconvolved using the default devoncolution algorithm in DDFacet (Tasse et al., in press).

\pg
The data was then time-averaged to create a new, 2.4 GB dataset with 1 data point per 8 seconds. Deliberately poor calibration was then performed on this dataset, solving for 1 calibration solution per 2 minutes ({\it caeteris paribus}). The resulting corrected residual data was imaged using the same imaging parameters as the reference image, and a cutout of the result is shown in Fig. \ref{image.3c295.nocorr}. As expected, the very long calibration intervals introduce calibration artefacts in the image. The brightest sources are still visible, but much of the fainter emission is buried under these artefacts. We are then in a case where our residual visibilities are dominated by calibration error rather than sky model incompleteness.

\pg
Weights were then calculated based on the badly-calibrated residual visibilities. Fig. \ref{image.3c295.lightcorr} was made using the same visibilities as Fig. \ref{image.3c295.nocorr} and applying baseline-based, sensitivity-optimal weight. Similarly, Fig. \ref{image.3c295.fullcorr} used the poorly-calibrated residual visibilities with the application of baseline-based, artefact-optimal weighting. These weights are likely to be poorly-conditioned. In both cases, all other parameters were conserved.

\pg
Note that applying antenna-based sensitivity-optimal weighting to the badly-calibrated data (not shown here) allows us to recover the reference image with only a very small increase in rms (increased by a factor of 1.14). Further testing on complex field simulations will be required to ascertain the usefulness of artefact-optimal weighting: it is likely that it fails to correct the image fully due to the poor conditioning of the covariance matrix used here.

\pg
{The pixel histograms show us that the weights do not completely mitigate the poor calibration interval choice, but certainly give a dramatic improvement over the unweighted, poorly-calibrated residuals. This is compatible with our statement that the weights give similar residuals in the image with a dramatic improvement in time at some cost in sensitivity. It is interesting to note that while Fig. \ref{image.3c295.fullcorr} looks noisier than Fig \ref{image.3c295.lightcorr}, its residual flux histogram is actually closer to that of Fig. \ref{image.3c295.goodcal}.}

\pg
As for performance, the weights used for Fig. \ref*{image.3c295.fullcorr} took 8 hours of computing time on a single core\footnote{Core type: Intel(R) Xeon(R) CPU E5-2660 0 @ 2.20GHz}, working on a 29 GB dataset, which is not particularly large for LOFAR data. Since the problem is massively parallel, this cost can be alleviated. The main bottleneck is likely due to very poor code optimization. As for the weights used for Fig. \ref*{image.3c295.antcorr}, they are computed in 1min6s on the same single core. 

%% file: source/discussion.tex
\section{Discussion}

This paper began by investigating the use of an algorithm inspired by ``lucky imaging'' to improve images made using radio interferometric data. By investigating the statistics of residual visibilities, we have made the following findings:

\begin{itemize}
\item A relationship between the statistics of residual visibilities and residual pixel values (the ``Cov-Cov relationship'').
\item A description of the noise-map in the image plane {as a constant variance level modulated by a noise-PSF convolved with the sources in the field. This gives }the variance in the flux of the image as a function of distance from the sources in the sky for a given calibration.
\item {An Adaptive Quality-Based weighting scheme, which reduces the noise in the image (and the presence of calibration artefacts) by minimising either the constant noise term or the noise-PSF.} 
\end{itemize}

\pg
{While our results are not a panacea for poor calibration, they show that we can not only improve images made with well-calibrated data, but also mitigate the worst effects of poorly-calibrated visibilities in otherwise well-calibrated datasets. Provided that the gain variability timescale is long enough at certain points of the observation, we can effectively get images of similar quality using both the} ``standard'' best-practice calibration interval for LOFAR survey data (calibration solution interval of 8 seconds) and a significantly larger solution interval of 2 minutes (frequency interval unchanged). {Of course, if no such stable interval exists, there will be no good intervals to upweigh, and we will be left only with equally-poor data chunks.} This means that{, in the right conditions,} net pipeline time {can be} sped up by a factor of nearly three, at a slight cost in sensitivity. {This increase will be greater than what could be achieved with existing comparable methods such as ``clipping''.} 

\pg
We emphasize that the adaptive quality-based weighting schemes work because the noise-{map} describes the \emph{underlying noise distribution}, of which calibration artefacts are \emph{one single realisation}. To fully characterise the artefacts, the correlation between different pixels (i.e. off-diagonal elements of $\Cov{\muely}$) must be computed; this has not been done in this paper. Nevertheless, lesser constraints on the spatial distribution of artefacts can be found using only the diagonal elements of $\Cov{\muely}$. The weighting schemes merely seek to {minimise} this spatial distribution as much as possible: the end result is \emph{fewer artefacts}, {which can be} distributed across a \emph{much larger area}. {This} is the source of the dramatic improvement from \ref{image.3c295.nocorr} to Fig. \ref{image.3c295.lightcorr}. We have simply down-weighted those visibilities where spurious signal was introduced by the calibration solutions, and up-weighted those visibilities where such signal was lesser. Since this spurious signal is the source of calibration artefacts, downweighting the associated visibilities reduces it dramatically. {The poor improvement from Fig. \ref{image.3c295.nocorr} to \ref{image.3c295.fullcorr}  is likely due to limits in the conditioning of our estimation of the covariance matrix.}

\pg
The work presented here can be improved upon, notably by working on improving the conditioning of our covariance matrix estimate: {for real observations}, it is impossible to have more than one realization of each gain value for all antennas. By treating each visibility within a calibration {interval} as a realization of the true distribution, we can better estimate the covariance matrix per baseline, and thus reach a better estimate of the variance in the image-plane. Of course, in practice, we can never access to the true, underlying time-covariance matrix for each baseline. Significant hurdles remain:
\begin{itemize}
\item The impact of sky model incompleteness (since calibration requires a sky model) is ignored in this paper; we start by assuming that we have a complete sky model. In practice, of course, acquiring a complete sky model is often a key science goal in and of itself. The impact of this hypothesis therefore ought to be investigated in future work.
\item The conditioning of our covariance matrix estimation remains a concern. By using an antenna-based approach, we can improve conditioning by a factor of $n_{ant}$, but this is only one approach among many. Further work is needed to investigate which method, if any, proves optimal.
\end{itemize}